# Integrated strong reciprocity enables productive punishment and protective defection

Tatsuya Sasaki[1,*] and Satoshi Uchida[2,3]

[1] Department of Community Development, Koriyama Women's College, Koriyama, Japan
[2] Research Center for Ethi-Culture Studies, RINRI Institute, Tokyo, Japan
[3] High-tech Research Centre, Kokushikan University, Tokyo, Japan

* Correspondence: t.sasaki@koriyama-kgc.ac.jp

January 5, 2025

## Abstract

Cooperation in large groups and one-shot interactions is often hindered by freeloading. Punishment can enforce cooperation, but it is usually regarded as wasteful because the costs of punishing offset its benefits. Here, we analyze an evolutionary game model that integrates upstream and downstream reciprocity with costly punishment—integrated strong reciprocity (ISR). We demonstrate that ISR admits a stable mixed equilibrium of ISR and unconditional defection (ALLD), and that costly punishment can become productive: When sufficiently efficient, it raises collective welfare above the no-punishment baseline. ALLD players persist as evolutionary shields, preventing invasion by unconditional cooperation (ALLC) or alternative conditional strategies (e.g., antisocial punishment). At the same time, the mixed equilibrium of ISR and ALLD remains robust under modest complexity costs that destabilize other symmetric cooperative systems.

## Significance Statement

Punishment is widely regarded as a costly yet necessary tool for enforcing cooperation. We present a new model—integrated strong reciprocity (ISR)—in which punishment is not only stabilizing but also productive, raising group welfare by sustaining higher levels of cooperation. Surprisingly, defectors can stably coexist and play a constructive role: By absorbing competition, they shield cooperation from invasion by other freeloaders or antisocial behaviors. ISR also withstands modest complexity costs that destabilize simpler reciprocity systems. Our results show that coercion itself can become adaptive in evolutionary dynamics. Taken together, our results point to a simple message: productive punishment, protective defection, and adaptive coercion.

# 1 Introduction

Human cooperation among strangers often persists even in one-shot interactions and large, well-mixed populations, where theory predicts that freeloading should occur. Social sanctioning is a leading explanation, but its persistence presents a long-standing evolutionary puzzle (Rankin et al., 2009; Sigmund, 2007). Punishment seems wasteful, as the costs borne by punishers and the losses imposed on targets can reduce group welfare. As a result, standard theory treats punishment as inefficient and vulnerable to "second-order freeloaders," who cooperate while avoiding enforcement costs (Panchanathan & Boyd, 2004; Boyd et al., 2003). Classic analyses further conclude that, if punishment is ever efficient, it operates only within a narrow range of conditions, and that refusing help to ill-reputed individuals is usually more effective than imposing sanctions (Ohtsuki et al., 2009; Rand et al., 2009). Yet across cultures, punishment remains widespread, ranging from informal sanctions in small groups to formal legal systems (Henrich et al., 2006; Fehr & Gächter, 2002). This tension suggests that existing theories overlook a mechanism explaining why punishment is not only common but also stable.

Research on reciprocity (Trivers, 1971) offers essential clues. Upstream reciprocity ("pay-it-forward" helping) spreads cooperation through emotion-based responses rather than deliberate strategies (Nowak & Roch, 2007; Gray et al., 2014). Downstream reciprocity (reputation-based helping) channels cooperation toward well-reputed partners while avoiding defectors (Alexander, 1987; Milinski et al., 2002; Nowak & Sigmund, 2005). Recent models demonstrate that integrating these "positive" reciprocity mechanisms (Baker & Bulkley, 2014) can sustain cooperation even in one-shot settings, where reciprocators coexist with defectors in stable equilibria—without the need for punishment, repeated play, or network structure (Sasaki et al., 2024, 2025; see also Pal et al., 2024). This highlights strategic diversity as a stabilizing force. However, except for a few examples (Oya & Ohtsuki, 2017; Kim & Murase, 2025; Glynatsi et al., 2025), these frameworks stop short of examining how reciprocity interacts with punishment, or whether such integration could overcome the efficiency limits that undermine punishment in standard models. Indeed, natural selection likely shaped these mechanisms to work synergistically rather than in isolation, particularly given evidence that the same individuals who show strong reciprocity (SR; downstream reciprocity + peer punishment) (Gintis, 2000; Bowles & Gintis, 2004; Richerson & Boyd, 2005; Gächter et al., 2008) also engage in altruistic punishment (Fehr & Gintis, 2007; Watanabe et al., 2014). Nevertheless, to our knowledge, no theoretical framework has yet integrated all three mechanisms—upstream reciprocity, downstream reciprocity, and costly punishment—into a single evolutionary model.

We address these limits by studying *integrated strong reciprocity* (ISR)—a single conditional strategy that unifies three elements often treated separately: (i) upstream reciprocity (pay-it-forward helping after being helped), (ii) downstream reciprocity (reputation-based helping of well-reputed partners), and (iii) costly punishment of ill-reputed partners. Embedding ISR in a one-shot helping game with replicator dynamics allows us to revisit three unresolved questions in cooperation theory: First, can punishment generate genuine welfare gains (Han et al., 2025) rather than merely redistribute resources? Second, what role does strategic diversity play—can defectors themselves contribute functionally to sustaining cooperation? Third, how do cognitive requirements and their costs (Imhof et al., 2005; Suzuki & Kimura, 2013; Yamamoto et al., 2024) shape the stability of punishment-based systems?

We address these questions analytically. ISR generates a stable polymorphism in which ISR coexists with defectors, resulting in global bistability between full defection (ALLD) and the ISR–ALLD mixed state. Within this polymorphism, ISR's punishment can become *productive*: When sufficiently efficient, sanctions raise average welfare above the no-punishment baseline. Mechanistically, defectors persist not as parasites but as *evolutionary shields*, absorbing selective pressure that would otherwise favor second-order freeloaders or antisocial punishers (Herrmann et al., 2008; Rand et al., 2010; Hauser et al., 2014). Finally, small cognitive or implementation costs that collapse standard strong reciprocity (SR = ISR without the upstream component) do not undermine ISR's coexistence; instead, ISR remains structurally stable, clarifying that sanctioning can be both viable and efficiency-enhancing. In short: *productive punishment, protective defection, and adaptive coercion.*

# 2 Results

## 2.1 Game structure

We model cooperation using a one-shot helping game, a standard framework for analyzing reciprocity (Sigmund, 2010). Each round, a randomly paired donor can choose to help the recipient, incurring a cost $c > 0$ while conferring a benefit $b > c$ to the partner. If the donor refuses to help, there is no exchange of costs or benefits. Players also have binary reputations—"Good" or "Bad"—that are publicly observable. Reputation updates depend on the donor's action and the recipient's reputation, following social norms that specify whether helping or refusing is judged positively or negatively.

Within this setting, we examine three classes of strategies. Unconditional defectors (ALLD) never help. Unconditional cooperators (ALLC) always help. While ALLC represents a typical second-order freeloader, ALLD acts as a first-order freeloader. Integrated strong reciprocity (ISR) strategy combines three conditional rules: (i) upstream reciprocity—helping a new partner after having been helped in the previous round; (ii) downstream reciprocity—helping recipients with a Good reputation but not those with a Bad reputation; and (iii) punishment—imposing a penalty on Bad recipients. ISR therefore unifies mechanisms that are typically studied separately, embedding them into a single evolutionary strategy.

**Reputation assignment.** We employ a strict assessment rule that serves as a "multi-factor verification" social norm, proposed by Sasaki et al. (2025), to distinguish ISR from other strategies. An individual is assessed as "Good" if and only if they satisfy all the requirements of the ISR strategy: (i) Upstream Check: cooperating with co-players if the individual received help in the previous round; (ii) Downstream Check: cooperating selectively with "Good" co-players when no help was received previously; and (iii) Punishment Check: punishing "Bad" co-players under the same condition as (ii). Failing to comply with any of these results in a Bad reputation; otherwise, an individual maintains a Good reputation, provided as the initial condition. Assessment noise is omitted (or even suppressed by costly cognitive systems) here.

This rule naturally embeds second-order punishment: In a mixed population containing ALLDs, ALLC players, by indiscriminately helping Bad ALLDs, fail the Downstream Check and are consequently assessed as Bad, thus becoming targets of punishment themselves. In contrast, ISRs pass the Triple Checks and maintain a Good reputation. Without ALLDs in the population, ALLCs' reputations would stay Good.

Population dynamics follow the replicator equation (Hofbauer & Sigmund, 1998). Let $x$, $y$, and $z$ denote the frequencies of ALLC, ALLD, and ISR, respectively, with $x + y + z = 1$. Expected interactions under the reputation dynamics determine payoffs, and strategies with above-average payoffs increase in frequency. This framework allows us to compare ISR with ALLD and ALLC under consistent evolutionary assumptions.

## 2.2 Payoffs

To evaluate evolutionary dynamics, we derive the expected payoffs of each strategy.

- ISR donor: helps a recipient with a Good reputation, pays a cost $c > 0$ while giving a benefit $b > c$ to the recipient, and imposes punishment on Bad recipients. When punishing, the ISR donor incurs a cost $\gamma > 0$ while imposing a penalty $\beta > 0$ on the recipient. ISR donors also follow the principle of upstream reciprocity: If they have been previously helped, they extend help to their next partner, regardless of the partner's reputation.
- ALLC donor: always helps, giving $b$ at a cost $c$ each time but never punishing.
- ALLD donor: never helps and never punishes, thereby avoiding all costs but also forgoing cooperative benefits.

The payoff of a strategy is given by the expected net benefit across all possible encounters, weighted by the current population frequencies. The expected payoff of each strategy, denoted $P_i$ with strategy $i$, can be described as follows:

$$P_{\text{ISR}} = \text{benefits received from others} - \text{costs of conditional helps}$$
$$- \text{costs of conditional punishments,}$$
$$P_{\text{ALLC}} = \text{benefits received from others} - \text{costs of unconditional helps}$$
$$- \text{penalties from conditional punishments,}$$
$$P_{\text{ALLD}} = \text{benefits received from others} - \text{penalties from conditional punishments.}$$

Exact expressions depend on the fraction of individuals judged. See Supplementary Information (SI) for details.

## 2.3 Evolutionary dynamics

Because ISR combines positive reciprocity with punishment, its relative performance depends on the helping-game efficiency ($b, c$) and punishment efficiency ($\beta, \gamma$). An analytical comparison of payoffs reveals when mixed equilibria of ISR and ALLD emerge. In particular, ISR and ALLD frequently coexist, whereas ISR in isolation is generally unstable, as it is susceptible to indirect invasions initiated by neutral drift from ALLC (van Veelen, 2011). This coexistence is central to our later results: It creates conditions in which punishment can be welfare-enhancing and in which defectors contribute functionally to the stability of cooperation.

Two equilibria generically arise when ISR and ALLD coexist. The condition $P_{\text{ISR}} = P_{\text{ALLD}}$ provides a quadratic equation of $z$. This leads to up to two roots in the interval of $(0,1)$, denoted $z_1$ and $z_2$ (with $z_1 < z_2$), corresponding to unstable and stable rest points ($P_1$ and $P_2$ in Fig. 1B), respectively. The stable equilibrium $z_2$ represents the long-run coexistence of ISR and ALLD, while $z_1$ marks the threshold below which ISR cannot invade a defector-dominated population. Thus, the global dynamics are *bistable*: The system converges either to full defection or to coexistence of ISR and defectors, depending on the initial conditions, as shown in Fig. 1B.

This mixed equilibrium is the cornerstone of our analysis. It is here that punishment can become *productive*, raising group welfare above the no-punishment baseline, and where defectors paradoxically stabilize cooperation rather than undermining it.

## 2.4 Welfare analysis

A central question is whether punishment merely redistributes resources between punishers and defectors, or whether it can generate genuine welfare gains at the population level (Han et al., 2025). We address this by comparing the average payoff at the mixed equilibrium with the corresponding payoff in the no-punishment case (Fig. 2).

Let $\bar{P}$ denote the population's average payoff at an equilibrium. Without punishment ($\beta = \gamma = 0$), cooperation can be sustained only through positive reciprocity. In this case, the unique interior equilibrium emerges at $z_0 = (b-2c)/(b-c)$ ($P_0$ in Fig. 1A), corresponding to the coexistence of integrated indirect reciprocators (IIR; Sasaki et al., 2024, 2025) and ALLD. With punishment active ($\beta, \gamma > 0$), the mixed equilibrium shifts to $z_2$ (with $z_0 < z_2$), where ISR coexists with ALLD (Fig. 1B).

An analytical comparison shows that punishment is *productive*:

$$\bar{P}(z_2) > \bar{P}(z_0). \qquad (1)$$

This inequality holds when the benefit-to-cost ratio $b/c$ is sufficiently high and punishment efficiency $\beta/\gamma$ exceeds a threshold. Intuitively, punishments deter excessive defection, thereby increasing the equilibrium frequency of ISR and raising the total amount of cooperation.

The condition for productive punishment can be understood in terms of a balance between benefits and costs. If punishment is too weak or too costly, the losses borne by punishers outweigh the cooperative gains, leaving group welfare unchanged or reduced. However, when punishment is sufficiently efficient, the cooperative benefits it enables more than compensate for the enforcement costs. In this regime, punishments cease to be a wasteful deterrent and become a net contributor to welfare.

## 2.5 Stability under cognitive costs

So far, we have assumed that strategies carry no cognitive or implementation costs. In practice, strategies that require monitoring emotions and reputations and remembering past interactions may impose additional burdens (Sasaki et al., 2025). To capture this, we introduce a small complexity cost $d > 0$ applied to conditional strategies (Imhof et al., 2005; Suzuki & Kimura, 2013; Yamamoto et al., 2024).

We first consider standard strong reciprocity (SR), which combines downstream reciprocity with punishment but lacks the upstream component of ISR (Fig. 1C). Under SR, even a small cost $d$ destabilizes cooperation: SR is outcompeted by ALLC, and a continuum of mixed equilibria with SR and ALLC vanishes. This reproduces the familiar problem that punishment alone, though potentially effective in the short term, cannot be sustained once its cognitive demands or implementation difficulties are recognized (Imhof et al., 2005; Suzuki & Kimura, 2013; Yamamoto et al., 2024).

By contrast, integrated strong reciprocity (ISR) remains robust under modest costs. Because ISR combines upstream and downstream reciprocity with punishment, defectors continue to serve as *evolutionary shields*, protecting ISR from invasion by alternative conditional strategies. As a result, the mixed equilibrium $z_2$ can persist even when $d > 0$. In other words, ISR can remain stable in environments where SR collapses.

This comparison underscores the importance of integration. Upstream reciprocity not only broadens the scope of cooperation but also enhances the evolutionary resilience of punishment-based systems. By embedding punishment within a larger reciprocity framework, ISR clarifies how costly punishment can remain viable despite modest cognitive costs.

# 3 Discussion

This study establishes two central results that directly address the three guiding questions posed in the Introduction. First, punishment can yield genuine welfare gains when embedded in the ISR framework, answering the first question. Second, strategic diversity can play a constructive role in sustaining cooperation, addressing the second question. Third, the cognitive requirements and their costs shape the stability of reciprocity systems differently under ISR and SR, providing insight into the third question. Together, these findings reframe costly punishment from a wasteful deterrent into a mechanism that can both stabilize cooperation and enhance welfare.

Our analysis identifies the conditions under which costly punishment becomes productive: when the cooperative gains it sustains exceed the costs of enforcement. This directly answers our first question. In this regime, punishments are not merely redistributive but raise the group's average welfare. The result parallels what is described as a facultative response to punishments by cooperating (Gardner & West, 2004; Lehmann & Keller, 2006; Roberts, 2013). In previous models, individuals strategically increase cooperation in the presence of frequent punishers, producing a positive correlation between punishment and prosocial behavior. In contrast, our framework does not rely on such behavioral flexibility: Natural selection acting on ISR alone produces the same functional outcome. In this sense, punishments are not only coercive but also investments in cooperation, stabilizing prosocial behavior and enhancing collective welfare.

Our results also extend Hardin's (1968) principle of "mutual coercion, mutually agreed upon" beyond deliberate institutional design into evolutionary dynamics. Hardin argued that communities must consciously accept coercive rules that restrict individual freedom to sustainably manage common resources. In our ISR framework, an analogous mechanism emerges spontaneously at the mixed equilibrium: ISRs impose fines on ALLDs through punishments, while ALLDs accept these fines as the implicit "price" of participating in cooperative societies. Crucially, both achieve higher payoffs at the asymmetric equilibrium than under costly-punishment monomorphism, which is vulnerable to ALLCs, second-order freeloaders, leading to symmetric defection. In this sense, mutual coercion does not necessarily depend on explicit agreement; it can self-organize through natural selection when punishment is sufficiently efficient. ISR thus broadens Hardin's insight, showing that coercion itself can become an adaptive feature of evolutionary dynamics.

Productive punishment has two noteworthy implications. First, it clarifies why a punitive clause can endure in human groups without centralized authority. When embedded in integrated reciprocity, punishment becomes part of an efficiency-enhancing equilibrium rather than a deadweight loss. Second, it explains the recurrent observation that a minority of non-cooperators can persist even in cooperative communities without undermining the performance of the whole. In our model, this minority is not a failure mode but a structural component that helps deter the rise of second-order freeloaders.

A distinctive feature of ISR is that defectors do not merely undermine cooperation but can serve as evolutionary shields. This finding directly addresses our second question about the role of strategic diversity. In many evolutionary models, cooperation is stabilized by eliminating selfish and antisocial behaviors. In contrast, ISR supports stable polymorphisms in which first-order freeloaders persist while functionally protecting reciprocity from erosion by second-order freeloaders or antisocial punishers. This perspective aligns with broader theoretical work emphasizing that cooperation often depends on the coexistence of heterogeneous strategies rather than the dominance of a single type (e.g., Nowak & Sigmund, 1998; Szabó & Fáth, 2007; Rand & Nowak, 2013). In this light, defectors are not merely tolerated but become integral to the architecture of stability: Their presence enables ISR to remain evolutionarily viable and, under productive conditions, to enhance collective welfare.

Finally, our framework clarifies how cognitive requirements shape the stability of punitive systems, addressing the third question. Standard strong reciprocity (SR), which combines reputation-based helping with punishment but lacks upstream reciprocity, collapses when even modest cognitive or implementation costs are introduced. By contrast, ISR remains robust: Upstream reciprocity broadens the pathways by which cooperation is rewarded, while ALLD continues to act as an evolutionary shield that protects ISR from displacement. This integration allows costly punishment to persist as a viable strategy even in environments where the burden of monitoring and memory would otherwise erode cooperative systems. In this way, ISR demonstrates how embedding punishment within a broader reciprocity framework can reconcile the benefits of costly punishment with realistic constraints on cognition.

Our analysis also has limitations that suggest promising directions for future research. We studied infinite, well-mixed populations with binary reputations and public assessment, focusing on a one-shot helping game. Extending the ISR model to finite populations (Schmid et al., 2021), structured populations (Murase & Hilbe, 2024), coordinated peer punishment (Boyd et al., 2010; García & Traulsen, 2019), richer reputation systems (Ohtsuki & Iwasa, 2006; Santos et al., 2018; Okada, 2020), ternary reputations (Yamamoto et al., 2025), private assessment (Uchida, 2010; Michel-Mata et al., 2024), stochastic errors (Brandt & Sigmund, 2006), partner choice (Graser et al., 2025), or other multiplayer dilemmas (Suzuki & Akiyama, 2005), such as public goods games (Sigmund et al., 2010; Li & Mifune, 2023; Sasaki et al., 2025), are natural next steps. Such extensions will test the generalizability of our findings and clarify how productive punishment and protective defection operate in more realistic settings.

In summary, integrated strong reciprocity reveals how reciprocity, defection, and punishment can interact to create cooperative systems that are both stable and welfare-enhancing. Costly punishment becomes productive when it raises group welfare above the no-punishment baseline. All-out defectors, often cast as obstacles (Axelrod & Hamilton, 1981), can serve as protective elements that shield reciprocity from collapse. By embedding peer punishment within a broader reciprocity framework, ISR enables adaptive coercion that remains viable even in the face of complexity costs. Together, these results challenge the view of costly punishment as inherently wasteful and reframe it as a cornerstone of evolutionary cooperation. In short: *productive punishment, protective defection, and adaptive coercion*.

## 4 Materials and methods

We study an infinite, well-mixed population repeatedly engaged in a one-shot helping game. In each encounter, a donor can help a recipient, providing benefit $b > 0$ at cost $c > 0$. Because $b > c$, helping is individually costly but socially beneficial.

**Strategies and reputation.** The population consists of three pure strategies: ALLC (always cooperate), ALLD (always defect), and ISR (integrated strong reciprocity). ISR combines upstream reciprocity (helping when helped in the previous round), downstream reciprocity (helping a Good player even when not helped in the previous round), and peer punishment (penalizing a Bad player at a cost to oneself).

We employ a binary reputation system with labels "Good" and "Bad." In brief, in a mixed population containing ALLD, the reputation assignment is as follows: Players are classified as Good if and only if they adhere strictly to the ISR strategy; otherwise, they are classified as Bad. This is equivalent to applying the action-based criteria (or "Triple Checks") defined in Section 2.

We further assume that players have access to sufficient information—through direct observation, records of past interactions, or indirect channels such as gossip—so that ISR players can reliably distinguish whom to punish (Bad) and whom not to punish (Good). Assessment noise is not incorporated into the main analysis; robustness to small noise can be analyzed analogously.

**Payoffs.** Let $(x, y, z)$ denote the frequencies of ALLC, ALLD, and ISR, respectively, with $x + y + z = 1$. The expected payoff of strategy $i$ is denoted $P_i$, and the average payoff of the whole population is given by $\bar{P} = xP_{\text{ISR}} + yP_{\text{ALLD}} + zP_{\text{ALLC}}$.

The expected payoffs for the three strategies are

$$\begin{aligned}
P_{\text{ALLC}} &= b(x + z(x + z)) - c - \beta z y, \\
P_{\text{ALLD}} &= b(x + z(x + z)) - \beta z y, \\
P_{\text{ISR}} &= b(x + z) - c((x + y)(x + z) + z) - \gamma(x + y)y - d,
\end{aligned} \qquad (2)$$

where $\beta > 0$ is the penalty imposed on punishees, $\gamma > 0$ the punisher's cost, and $d > 0$ the complexity cost.

Equation (2) first results in the fact that ALLD is always better off with $c > 0$ than ALLC is and thus that there exists no interior equilibrium in the state space.

**Evolutionary dynamics.** We analyze the evolutionary dynamics of ISR, ALLD, and ALLC using the replicator equation. The replicator equation specifies how strategy frequencies change over time:

$$\dot{x} = x(P_{ALLC} - \bar{P}), \qquad \dot{y} = y(P_{ALLD} - \bar{P}), \qquad \dot{z} = z(P_{ISR} - \bar{P}). \qquad (3)$$

Strategies with above-average payoffs increase in frequency, while those with below-average payoffs decline. This framework ensures that the population evolves toward equilibria where strategies earn equal payoffs, or toward boundary states occupied by fewer strategies.

On the ISR–ALLD subspace with $x = 0$, the equation system in Eq. (3) reduces to

$$\dot{z} = z(1-z)G(z), \qquad (4)$$

with

$$G(z) \coloneqq P_{ISR} - P_{ALLD} = -(b - c + \beta + \gamma)z^2 + (b - 2c + \beta + 2\gamma)z - (d + \gamma). \qquad (5)$$

Because $G(z)$ is concave, it admits up to two internal roots $0 < z_1 < z_2 < 1$. The larger root $z_2$, if it exists, is an asymptotically stable, mixed state of ISR and ALLD.

It is evident that on the ALLC–ISR subspace with $y = 0$, ALLC is always better off with $d > 0$ than ISR is. Therefore, (i) if $G(z)$ has different real roots, the whole system is bistable: all interior trajectories converge either to the mixed equilibrium with $z = z_2$ or to the uniform equilibrium with $z = 0$ (ALLD corner); and (ii) if $G(z)$ has no real roots, the whole system has the global attractor with $z = 0$.

**Collective welfare.** Population welfare is measured by the average payoff $\bar{P}$ at equilibrium. Of special interest is whether costly punishment is *productive*: that is, whether equilibria with costly punishment yield higher $\bar{P}$ than those without. We first consider the case with $d = 0$.

When $\beta = \gamma = 0$ and $b > 2c$, ISR reduces to integrated indirect reciprocity (IIR), and Eq. (5) degenerates to a linear function with the unique interior root, $0 < z_0 = (b - 2c)/(b - c) < 1$ (Sasaki et al., 2025). The population average payoff at $z = z_0$ is given by

$$\bar{P}(z_0) = bz_0^2. \qquad (6)$$

Introducing costly punishment with $\beta, \gamma > 0$ yields the new stable equilibrium $z = z_2$, at which point,

$$\bar{P}(z_2) = bz_2^2 - \beta z_2(1 - z_2). \qquad (7)$$

Analytically, one can show that $\frac{dz_2}{d\beta} > 0$: Higher penalties increase the equilibrium share of ISR. Moreover, the derivative of welfare satisfies

$$\frac{\partial \bar{P}(z_2)}{\partial \beta} = \frac{\partial z_2}{\partial \beta} \cdot H(z_2), \qquad (8)$$

with

$$H(z_2) = 2cz_2 + (b - 2c) + 2\gamma(1 - z_2). \qquad (9)$$

This expression is positive under broad conditions—for example, whenever $b \geq 2c$—so that $\bar{P}(z_2)$ increases with $\beta$. In these regimes, costly punishment is unequivocally *productive*, expanding cooperation and enhancing welfare.

**Note:** Full derivations—including conditions for internal equilibria, transversal stability, Jacobian analyses, and welfare inequalities—are presented in the Supplementary Information (SI).

**Conflict of Interest**

The authors declare that the research was conducted in the absence of any commercial or financial relationships that could be construed as a potential conflict of interest.

**Author Contributions**

TS: Visualization, Conceptualization, Formal analysis, Methodology, Writing – original draft, Writing – review & editing, Investigation. SU: Conceptualization, Methodology, Writing – review & editing.

**Funding**

This work was supported by JSPS KAKENHI, Grant Numbers 23K05943 (TS).

# References

1. Rankin DJ, Dos Santos M, Wedekind C (2009) The evolutionary significance of costly punishment is still to be demonstrated. Proc Natl Acad Sci USA 106:E135.
2. Sigmund K (2007) Punish or perish? Retaliation and collaboration among humans. Trends Ecol Evol 22:593–600.
3. Panchanathan K, Boyd R (2004) Indirect reciprocity can stabilize cooperation without the second-order free rider problem. Nature 432:499–502.
4. Boyd R, Gintis H, Bowles S, Richerson PJ (2003) The evolution of altruistic punishment. Proc Natl Acad Sci USA 100:3531–3535.
5. Ohtsuki H, Iwasa Y, Nowak MA (2009) Indirect reciprocity provides only a narrow margin of efficiency for costly punishment. Nature 457:79–82.
6. Rand DG, Ohtsuki H, Nowak MA (2009) Direct reciprocity with costly punishment: Generous tit-for-tat prevails. J Theor Biol 256:45–57.
7. Henrich J, et al. (2006) Costly punishment across human societies. Science 312:1767–1770.
8. Fehr E, Gächter S (2002) Altruistic punishment in humans. Nature 415:137–140.
9. Trivers RL (1971) The evolution of reciprocal altruism. Q Rev Biol 46:35–57.
10. Nowak MA, Roch S (2007) Upstream reciprocity and the evolution of gratitude. Proc R Soc B 274:605–610.
11. Gray K, Ward AF, Norton MI (2014) Paying it forward: Generalized reciprocity and the limits of generosity. J Exp Psychol Gen 143:247–254.
12. Alexander RD (1987) The Biology of Moral Systems (Aldine de Gruyter, New York).

13. Milinski M, Semmann D, Krambeck HJ (2002) Reputation helps solve the 'tragedy of the commons'. Nature 415:424–426.
14. Nowak MA, Sigmund K (2005) Evolution of indirect reciprocity. Nature 437:1291–1298.
15. Baker WE, Bulkley N (2014) Paying it forward vs. rewarding reputation: Mechanisms of generalized reciprocity. Organ Sci 25:1493–1510.
16. Sasaki T, Uchida S, Okada I, Yamamoto H (2024) The evolution of cooperation and diversity under integrated indirect reciprocity. Games 15:15.
17. Sasaki T, Uchida S, Okada I, Yamamoto H, Nakai Y (2025) Integrating upstream and downstream reciprocity stabilizes cooperator-defector coexistence in N-player giving games. arXiv [Preprint] 2509.04743.
18. Pal S, Hilbe C, Glynatsi NE (2024) The co-evolution of direct, indirect and generalized reciprocity. arXiv [Preprint] 2411.03488.
19. Oya G, Ohtsuki H (2017) Stable polymorphism of cooperators and punishers in a public goods game. J Theor Biol 419:243–253.
20. Kim H, Murase Y (2025) Incomplete reputation information and punishment in indirect reciprocity. arXiv [Preprint] 2509.09181.
21. Glynatsi NE, Hilbe C, Murase Y (2025) Exact conditions for evolutionary stability in indirect reciprocity under noise. PLoS Comput Biol 21:e1013584.
22. Gintis H (2000) Strong reciprocity and human sociality. J Theor Biol 206:169–179.
23. Bowles S, Gintis H (2004) The evolution of strong reciprocity: cooperation in heterogeneous populations. Theor Popul Biol 65:17–28.
24. Richerson PJ, Boyd R (2005) Not by Genes Alone: How Culture Transformed Human Evolution (Univ of Chicago Press, Chicago).
25. Gächter S, Renner E, Sefton M (2008) The long-run benefits of punishment. Science 322:1510.
26. Fehr E, Gintis H (2007) Human motivation and social cooperation: Experimental and analytical foundations. Annu Rev Sociol 33:43–64.
27. Watanabe T, et al. (2014) Two distinct neural mechanisms underlying indirect reciprocity. Proc Natl Acad Sci USA 111:3990–3995.
28. Han TA, Duong MH, Perc M (2024) Evolutionary mechanisms that promote cooperation may not promote social welfare. J R Soc Interface 21:20240547.
29. Imhof LA, Fudenberg D, Nowak MA (2005) Evolutionary cycles of cooperation and defection. Proc Natl Acad Sci USA 102:10797–10800.
30. Suzuki S, Kimura K (2013) Indirect reciprocity is sensitive to costs of information transfer. Sci Rep 3:1435.
31. Yamamoto H, Okada I, Sasaki T, Uchida S (2024) Clarifying social norms which have robustness against reputation costs and defector invasion in indirect reciprocity. Sci Rep 14:25073.
32. Herrmann B, Thöni C, Gächter S (2008) Antisocial punishment across societies. Science 319:1362–1367.
33. Rand DG, Armao JJ, Nakamaru M, Ohtsuki H (2010) Anti-social punishment can prevent the co-evolution of punishment and cooperation. J Theor Biol 265:624–632.
34. Hauser OP, Nowak MA, Rand DG (2014) Punishment does not promote cooperation under exploration dynamics when anti-social punishment is possible. J Theor Biol 360:163–171.
35. Sigmund K (2010) The Calculus of Selfishness (Princeton Univ Press, Princeton, NJ).
36. Hofbauer J, Sigmund K (1998) Evolutionary Games and Population Dynamics (Cambridge Univ Press, Cambridge, UK).

37. van Veelen M (2011) Robustness against indirect invasions. Games Econ Behav 74:382–393.
38. Gardner A, West SA (2004) Cooperation and punishment, especially in humans. Am Nat 164:753–764.
39. Lehmann L, Keller L (2006) The evolution of cooperation and altruism–a general framework and a classification of models. J Evol Biol 19:1365–1376.
40. Roberts G (2013) When punishment pays. PLoS One 8:e57378.
41. Hardin G (1968) The tragedy of the commons. Science 162:1243–1248.
42. Nowak MA, Sigmund K (1998) Evolution of indirect reciprocity by image scoring. Nature 393:573–577.
43. Szabó G, Fáth G (2007) Evolutionary games on graphs. Phys Rep 446:97–216.
44. Rand DG, Nowak MA (2013) Human cooperation. Trends Cogn Sci 17:413–425.
45. Schmid L, Chatterjee K, Hilbe C, Nowak MA (2021) A unified framework of direct and indirect reciprocity. Nat Hum Behav 5:1292–1302.
46. Murase Y, Hilbe C (2024) Computational evolution of social norms in well-mixed and group-structured populations. Proc Natl Acad Sci USA 121:e2406885121.
47. Boyd R, Gintis H, Bowles S (2010) Coordinated punishment of defectors sustains cooperation and can proliferate when rare. Science 328:617–620.
48. García J, Traulsen A (2019) Evolution of coordinated punishment to enforce cooperation from an unbiased strategy space. J R Soc Interface 16:20190127.
49. Ohtsuki H, Iwasa Y (2006) The leading eight: Social norms that can maintain cooperation by indirect reciprocity. J Theor Biol 239:435–444.
50. Santos FP, Santos FC, Pacheco JM (2018) Social norm complexity and past reputations in the evolution of cooperation. Nature 555:242–245.
51. Okada I (2020) A review of theoretical studies on indirect reciprocity. Games 11:27.
52. Yamamoto H, Okada I, Suzuki T (2025) Gradual reputation dynamics evolve and sustain cooperation in indirect reciprocity. PLoS One 20:e0329742.
53. Uchida S (2010) Effect of private information on indirect reciprocity. Phys Rev E 82:036111.
54. Michel-Mata S, Kawakatsu M, Sartini J, Kessinger TA, Plotkin JB, Tarnita CE (2024) The evolution of private reputations in information-abundant landscapes. Nature 634:883–889.
55. Brandt H, Sigmund K (2006) The good, the bad and the discriminator - Errors in direct and indirect reciprocity. J Theor Biol 239:183–194.
56. Graser C, Fujiwara-Greve T, García J, Van Veelen M (2025) Repeated games with partner choice. PLoS Comput Biol 21:e1012810.
57. Suzuki S, Akiyama E (2005) Reputation and the evolution of cooperation in sizable groups. Proc R Soc B 272:1373–1377.
58. Sigmund K, De Silva H, Traulsen A, Hauert C (2010) Social learning promotes institutions for governing the commons. Nature 466:861–863.
59. Li Y, Mifune N (2023) Punishment in the public goods game is evaluated negatively irrespective of non-cooperators' motivation. Front Psychol 14:1198797.
60. Axelrod R, Hamilton WD (1981) The evolution of cooperation. Science 211:1390–1396.

# Figures

## Figure 1

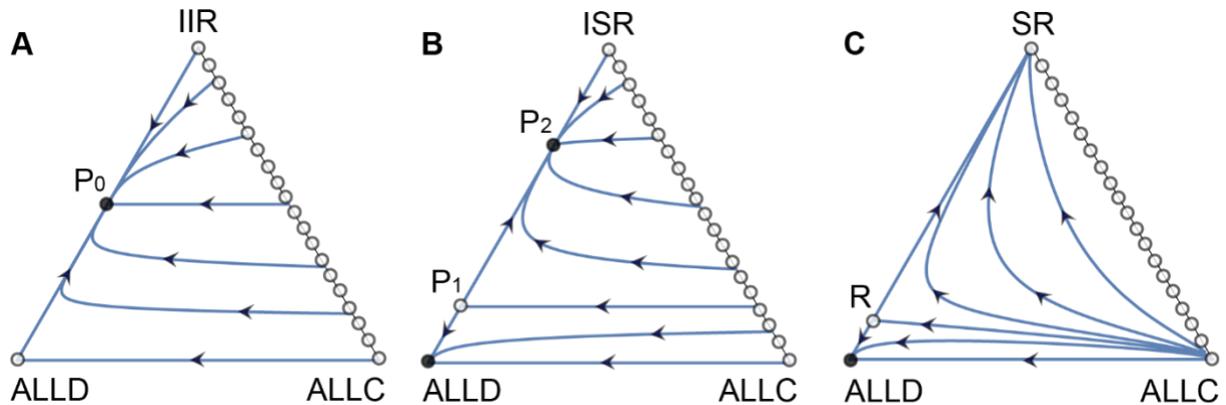

Figure 1. Evolutionary dynamics with no complexity costs. Phase portraits of the replicator dynamics for populations consisting of ALLC, ALLD, and a conditional strategist ($d = 0$). (**A**) Integrated Indirect Reciprocity (IIR). The system converges to a globally stable coexistence equilibrium $P_0$. (**B**) Integrated Strong Reciprocity (ISR). The dynamics exhibit bistability with an unstable invasion barrier $P_1$ and a stable mixed equilibrium $P_2$. Notably, the frequency of reciprocators at $P_2$ exceeds that at $P_0$ in Panel **A**, illustrating how punishment can increase the prevalence of cooperation (productive punishment). (**C**) Standard Strong Reciprocity (SR). The pure SR state is not asymptotically stable and is vulnerable to invasion by ALLC. The rest point R acts as a repeller; trajectories starting near SR eventually drift toward ALLC (neutral stability) and then collapse to ALLD, highlighting the lack of an evolutionary shield. Parameters: $b = 3$, $c = 1$, $\beta = 1.5$, $\gamma = 0.5$.

**Figure 2**

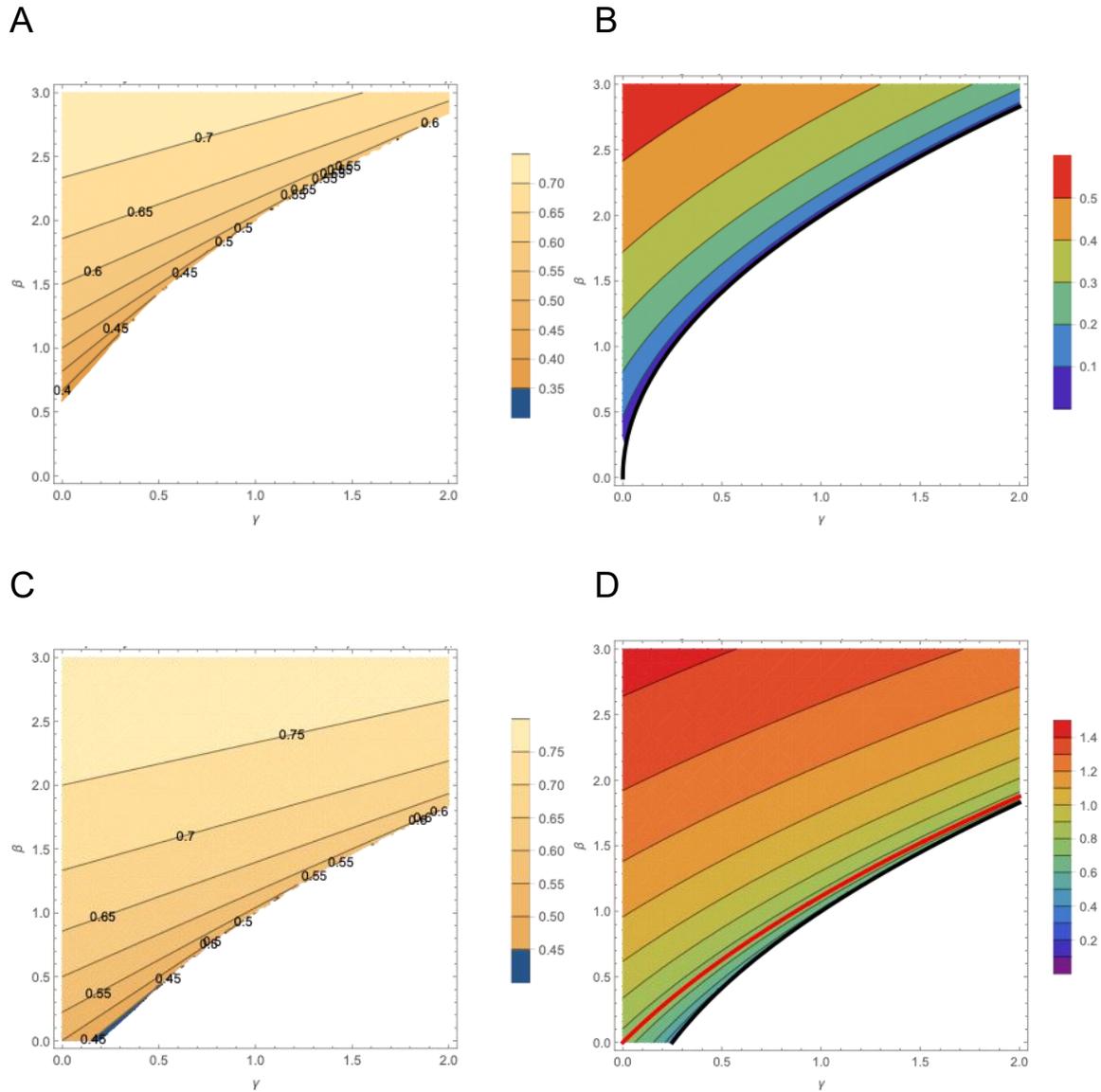

Figure 2. Welfare analysis with varying punishment parameters ($\gamma$, $\beta$). (**A**, **C**) Frequency of ISR at the stable mixed equilibrium of ISR and ALLD, (**B**, **D**) Average payoff at the stable mixed equilibrium of ISR and ALLD. In **C** and **D**, the stable mixed equilibrium is depicted as $P_2$ in Fig. 1**B**. In all panels, the blank region represents a region where the boundary dynamics along the $y$–$z$ edge have no mixed equilibrium of ISR and ALLD. In **D**, the red-colored curve depicts a curve along which the average payoff is 0.75, equally sufficient with that of no punishment in Eq. (6) (at the stable mixed equilibrium of IIR and ALLD, $P_0$ in Fig. 1**A**). Parameters: (**A**, **B**) $b = 2$, $c = 1$, and $d = 0$; (**C**, **D**) $b = 3$, $c = 1$, and $d = 0$.

# Supplementary Information

## S1 Model overview and strategy set

We consider an infinite, well-mixed population repeatedly playing a one-shot helping game. In each interaction of helping games considered, a randomly chosen donor decides whether to help a recipient, conferring a benefit $b > 0$ at a personal cost $c > 0$. We assume $b > c$, so that helping constitutes the Prisoner's Dilemma. In addition, when the recipient is judged "Bad" (see Reputation and assessment below), a "punishing" donor may impose a penalty $\beta > 0$ on the recipient at a personal cost $\gamma > 0$. Cognitive or implementation costs of conditional control are represented by $d \geq 0$ (introduced for ISR in §S6 and for SR in §S7).

The population state is given by the frequencies of three strategies: unconditional cooperators (ALLC), unconditional defectors (ALLD), and *integrated strong reciprocity* (ISR), denoted $(x, y, z)$ with $x + y + z = 1$ and $x, y, z \in [0,1]$.

**Strategies.**

- **ALLC**: helps every recipient and never punishes.
- **ALLD**: never helps and never punishes.
- **ISR**: integrates (i) **upstream reciprocity** (helps when the focal donor was helped in the previous round), (ii) **downstream reciprocity** (helps "Good" recipients when the focal donor was not helped in the previous round), and (iii) **peer punishment** (punishes "Bad" recipients when the focal donor was not helped in the previous round). ISR bears the complexity cost $d > 0$ when considered in §S6 (and $d = 0$ in baseline §§S3–S5).

**Reputation and assessment.** We adopt a binary reputation system (Good/Bad) with a discriminating ISR assessment rule. Roughly speaking, players whose actions are consistent with ISR are judged Good, and those that deviate from ISR are judged Bad.

**Payoffs and evolutionary dynamics.** We denote the expected payoffs of types as $P_{\text{ALLC}}$, $P_{\text{ALLD}}$, and $P_{\text{ISR}}$, respectively. Their explicit forms, which account for upstream reciprocity and punishment in ISR, are derived in §S3 (with $d = 0$) and extended in §§S6–S7 (with $d > 0$). The average payoff is

$$\bar{P} = xP_{\text{ALLC}} + yP_{\text{ALLD}} + zP_{\text{ISR}}. \qquad (S1)$$

We employ continuous-time replicator dynamics:

$$\dot{x} = x(P_{\text{ALLC}} - \bar{P}), \qquad \dot{y} = y(P_{\text{ALLD}} - \bar{P}), \qquad \dot{z} = z(P_{\text{ISR}} - \bar{P}). \qquad (S2)$$

Boundary dynamics on each edge (ALLC–ALLD, ALLD–ISR, ALLC–ISR) and interior equilibria are analyzed in §§S4–S7. In particular, we identify the coexistence geometry on the ALLD–ISR edge (a quadratic selection gradient with unstable and stable roots), assess welfare at attractors (§S5), and analyze the effect of ISR's complexity cost $d$ on both edge stability and transversal stability (§S6), alongside a parallel analysis for strong reciprocity (SR) (§S7).

## S2 Reputation dynamics

Players are assigned binary reputations—Good or Bad—based on their observed actions. We employ a strict assessment rule that serves as a "multi-factor verification" social norm, invented by Sasaki et al. (2025), to distinguish ISR from other strategies. An individual is assessed as "Good" if and only if they satisfy both of the following action-based criteria (an AND condition): (i) Upstream Check: cooperating with co-players if the individual received help in the previous round; otherwise (if the individual did not receive help), (ii) Downstream Check: cooperating selectively with "Good" co-players, and (iii) Punishment Check: punishing "Bad" co-players. Failing either condition results in a Bad reputation; otherwise, an individual maintains a Good reputation, provided as the initial condition. Assessment noise is omitted (or even suppressed by costly cognitive systems) here.

- **ISR**: always judged Good, since ISR's conditional behavior is aligned with the assessment rule.
- **ALLD**: always judged Bad, since ALLD never helps and never punishes.
- **ALLC**: in the population mixed with ALLD, judged Bad, since ALLC never punishes; in the population with no ALLD, judged Good, since rule deviations never happen.

## S3 Payoff functions

Based on the reputation assessment in §S2, we calculate the expected payoffs for each strategy. If a state is mixed with ALLD ($y > 0$), the probability that a random partner is Good is $\pi_G = z$, while the probability of a Bad partner is $\pi_B = x + y$; or otherwise ($y = 0$), $\pi_G = x + z = 1$.

### S3.1 Payoff of ALLD

An ALLD donor never helps and never punishes. Its expected payoff arises only from being helped by others:

- From ALLC ($x$): always helped.
- From ISR ($z$): helped by upstream reciprocity with probability $x + z$; punished when the ISR donor was not helped and the recipient is Bad, with probability $y$ (not helped) × 1 (recipient is Bad) = $y$.

Thus,

$$P_{\text{ALLD}}(x, y, z) = b[x + z(x + z)] - \beta z y. \quad (S3)$$

### S3.2 Payoff of ALLC

In states with ALLD, ALLC always helps and never punishes; its reputation becomes Bad.

- As donor: pays $c$ in every encounter.
- From ALLC ($x$): always helped.

- From ISR ($z$): helped by upstream reciprocity with probability $x + z$; punished when the ISR donor was not helped and the recipient is Bad, with probability $y$ (not helped) × 1 (recipient is Bad) = $y$.

Thus,

$$P_{\text{ALLC}}(x, y, z) = b[x + z(x + z)] - c - \beta zy. \qquad \text{(S4-1)}$$

In states with no ALLD ($y = 0$): always helped by both ALLC ($x$) and ISR ($z$), the latter motivated by upstream reciprocity.

Thus,

$$P_{\text{ALLC}}(x, y, z) = b - c. \qquad \text{(S4-2)}$$

### S3.3 Payoff of ISR (with upstream, downstream, and punishment)

**As donor.** ISR helps in two cases: (i) **Upstream reciprocity**—if the ISR donor was helped in the previous round (probability $x + z$), it helps regardless of the partner's reputation; (ii) Otherwise (probability $y$), **downstream reciprocity**—help only if the current recipient is Good (probability $\pi_G = z$). Hence, the ISR donor's helping probability is

$$h_{\text{ISR}} = (x + z) + yz. \qquad \text{(S5)}$$

ISR **punishes** only when it was not helped and meets a Bad recipient: probability $y \times (x + y)$. The expected donor costs are therefore $ch_{\text{ISR}}$ and $\gamma(x + y)y$.

**As recipient.** ISR is Good, so it is always helped by ALLC and—importantly—also helped by ISR: if the ISR donor was helped in the previous round, it helps by upstream reciprocity; if not, it still helps Good recipients by downstream reciprocity. Thus, an ISR recipient is helped by ISR with probability 1, yielding a total expected recipient benefit of $b(x + z)$.

Collecting these pieces, the ISR payoff is

$$P_{\text{ISR}} = b(x + z) - ch_{\text{ISR}} - \gamma h_{\text{ISR}} = b(x + z) - c[(x + z) + yz] - \gamma(x + y)y. \qquad \text{(S6)}$$

In §§S6–S7, we extend the analysis to include the complexity cost $d > 0$ for ISR and SR.

## S4 Boundary and global dynamics

### S4.1 On the $x$–$y$ edge (ALLC–ALLD, $z = 0$)

From Eqs. (S3)–(S4) with $z = 0$:

$$P_{\text{ALLD}} = bx, \qquad P_{\text{ALLC}} = bx - c. \qquad \text{(S7)}$$

Hence,

$$P_{\text{ALLD}} - P_{\text{ALLC}} = c > 0, \qquad \text{(S8)}$$

so ALLD strictly dominates ALLC. The *x–y* edge dynamics converge to the ALLD vertex ($y = 1$).

### S4.2 On the *y–z* edge (ALLD–ISR, $x = 0$)

From Eqs. (S3) and (S6) with $x = 0$:

$$P_{ALLD} = bz^2 - \beta zy, \qquad P_{ISR} = bz - c(2z - z^2) - \gamma y^2. \qquad (S9)$$

The replicator equation is

$$\dot{z} = z(1 - z)\, G(z) \quad \text{with} \quad G(z) := P_{ISR} - P_{ALLD}. \qquad (S10)$$

Using $y = 1 - z$ and simplifying:

$$G(z) = -(b - c + \beta + \gamma)z^2 + (b - 2c + \beta + 2\gamma)z - \gamma. \qquad (S11)$$

For compactness, define

$$A = b - 2c + \beta + 2\gamma, \qquad B = b - c + \beta + \gamma. \qquad (S12)$$

Note that $B > 0$, and thus, $G(z)$ is a concave quadratic function with (i) discriminant: $\Delta = A^2 - 4\gamma B$, (ii) vertex (maximum): $z^\star = \frac{A}{2B}$, and (iii) vertex value: $G(z^\star) = \frac{\Delta}{4B}$. Note $G(0) = -\gamma < 0$ and $G(1) = -c < 0$. Thus, $G(z)$ has two internal roots with $0 < z_1 < z_2 < 1$ if and only if

$$A > 0 \quad \text{and} \quad \Delta > 0. \qquad (S13)$$

Under Eq. (S13), the roots are

$$z_{1,2} = \frac{A \mp \sqrt{\Delta}}{2B}, \qquad (S14)$$

with $z_1$ unstable and $z_2$ stable (which both are mixed ISR–ALLD equilibria).

### S4.3 On the *x–z* edge (ALLC–ISR, $y = 0$)

From Eqs. (S4-2) and (S6) with $y = 0$, we have that $P_{ALLC} = P_{ISR} = b - c$, and thus the entire *x–z* edge is a neutral line.

### S4.4 Interior equilibria and global dynamics

For any interior state, it follows that

$$P_{ALLD} - P_{ALLC} = c > 0. \qquad (S15)$$

Thus, there exists no interior equilibrium; all interior trajectories converge to the *y–z* edge, either to the ALLD vertex or to the stable mixed equilibrium $(0, 1 - z_2, z_2)$.

# S5 Welfare conditions: average payoff and productive punishment

To evaluate whether punishment is *productive*—that is, whether it increases social welfare rather than merely redistributing payoffs—we examine the population's average payoff at stationary states.

## S5.1 Average payoff at the coexistence

At any coexistence point $\hat{z} \in \{z_1, z_2\}$, the equilibrium condition $G(\hat{z}) = 0$ ensures $P_{\text{ALLD}}(\hat{z}) = P_{\text{ISR}}(\hat{z})$. From Eq. (S9):

$$\bar{P}(\hat{z}) = P_{\text{ALLD}}(\hat{z}) = b(\hat{z})^2 - \beta \hat{z}(1 - \hat{z}). \quad (S16)$$

This decomposition is transparent: The first term represents the gains from cooperation, and the second term represents the welfare loss from punishment expenditures.

## S5.2 Benchmark: without punishment (IIR)

If $\beta = \gamma = 0$, ISR reduces to the "integrated indirect reciprocity" (IIR) strategy (Sasaki et al., 2024, 2025). Then, Eq. (S11) simplifies and yields a unique coexistence equilibrium at

$$z_0 = \frac{b - 2c}{b - c}, \quad \text{provided} \quad b > 2c. \quad (S17)$$

The corresponding average payoff is

$$\bar{P}(z_0) = b z_0^2. \quad (S18)$$

If $b \leq 2c$, IIR admits no coexistence equilibrium and ALLD dominates, so we define the baseline $\bar{P} = 0$.

## S5.3 Condition for productive punishment

We say punishment is *productive* if

$$\bar{P}(z_2) > \bar{P}(z_0). \quad (S19)$$

That is, the stable-coexistence payoff under ISR with $\beta, \gamma > 0$ exceeds the no-punishment benchmark.

The condition in Eq. (S19) is only relevant when the coexistence equilibrium exists, that is, when the *y–z* edge admits two internal roots.

## S5.4 Analytic sketch: dependence of $z_2$ and $\bar{P}(z_2)$ on $\beta$

**Step 1.** From Eq. (S11), we differentiate the frequency of stable root $z_2$ with respect to $\beta$. We find $\frac{\partial z_2}{\partial \beta}$ for $G(z_2) = 0$ by using implicit differentiation, resulting in

$$0 = \frac{\partial G(z_2)}{\partial \beta} = -z_2^2 + z_2 - \frac{\partial z_2}{\partial \beta}[2Bz_2 - A], \quad (S20)$$

thus,

$$\frac{\partial z_2}{\partial \beta} = \frac{-z_2^2 + z_2}{2Bz_2 - A}. \quad (S21)$$

At the stable root $z_2$, the slope satisfies $G'(z_2) = -2Bz_2 + A < 0$, since it corresponds to the right-hand root of a downward-opening quadratic. Therefore, the denominator is positive. Moreover, because $0 < z_2 < 1$, the numerator is positive, too. Hence,

$$\frac{\partial z_2}{\partial \beta} > 0. \quad (S22)$$

Intuitively, increasing $\beta$ lowers the relative payoff of ALLD, thereby shifting the right-hand intersection of $G(z)$ to the right (larger $z_2$). Thus, stronger punishment penalties monotonically increase the equilibrium share of ISR (Fig. 2A, 2C).

**Step 2.** Welfare at coexistence is $\bar{P}(z_2) = bz_2^2 - \beta z_2(1 - z_2)$. Differentiating with respect to $\beta$ gives

$$\frac{\partial \bar{P}(z_2)}{\partial \beta} = \frac{\partial z_2}{\partial \beta} \cdot H(z_2), \quad H(z) := (2bz - \beta(1 - 2z)) - (2Bz - A). \quad (S23)$$

Substituting $2Bz - A = 2(b - c + \beta + \gamma)z - (b - 2c + \beta + 2\gamma)$ into $H(z)$ leads to

$$H(z) = 2cz + (b - 2c) + 2\gamma(1 - z). \quad (S24)$$

**Corollary.** If $b \geq 2c$, then Eq. (S24) takes a positive value, and by Eq. (S22) we conclude

$$\frac{\partial \bar{P}(z_2)}{\partial \beta} > 0. \quad (S25)$$

## S5.5 Interpretation

- **Monotonicity.** Both the stable coexistence frequency of ISR $z_2$ and the corresponding welfare $\bar{P}(z_2)$ increase with the punishment penalty $\beta$ (Fig. 2).
- **Economic meaning.** Greater punishment intensity reduces the relative payoff of defectors, shifts the coexistence toward higher ISR prevalence, and thereby enhances collective welfare.
- **Productive punishment.** When $b \geq 2c$, Eq. (S25) guarantees that welfare rises with $\beta$ across the coexistence region. When $b < 2c$, IIR collapses to the 100%-ALLD state, in which $\bar{P} = 0$, but ISR may still generate positive welfare if punishment is sufficiently effective.

# S6 Complexity costs in ISR

We now examine the impact of complexity costs on ISR. Suppose that ISR players pay a small complexity cost $d > 0$ for executing their conditional rules. This modifies the payoff of ISR uniformly: $P_{ISR} \to P_{ISR} - d$.

## S6.1 Boundary dynamics with $d > 0$

**On the y–z edge ($x = 0$).** Substituting into Eq. (S11) gives

$$G_d(z) = -Bz^2 + Az - (\gamma + d). \qquad (S26)$$

The discriminant becomes $\Delta_d = A^2 - 4B(\gamma + d)$. If $A > 0$ and $\Delta_d > 0$, two internal roots exist:

$$z_{1,2}(d) = \frac{A \mp \sqrt{\Delta_d}}{2B}, \qquad 0 < z_1(d) < z_2(d) < 1. \qquad (S27)$$

As before, $z_1(d)$ is unstable and $z_2(d)$ is stable. Notably,

$$\frac{\partial z_2}{\partial d} < 0, \qquad (S28)$$

so the stable coexistence point $z_2$ shifts toward the ALLD vertex as $d$ increases.

**On the x–z edge ($y = 0$).** Differentiating the payoffs gives

$$P_{ALLC} - P_{ISR} = d > 0. \qquad (S29)$$

Thus, with any positive cognitive cost, ALLC strictly dominates ISR along the x–z edge. The neutrality present at $d = 0$ is destroyed.

**On the x–y edge ($z = 0$).** ALLD continues to dominate ALLC: $P_{ALLD} - P_{ALLC} = c > 0$.

## S6.2 Global dynamics with $d > 0$

ALLD dominates ALLC no matter what the complexity cost, $d$. Thus, at the mixed equilibrium point $(0, 1 - z_2(d), z_2(d))$, ALLC cannot invade the point, ensuring **transversal stability** of the boundary equilibrium.

Therefore, ISR cannot survive in isolation against ALLC, but ISR can coexist with ALLD. The global phase portrait remains **bistable**: Trajectories converge either to the 100%-ALLD state or to the ISR–ALLD mixed equilibrium.

## S6.3 Why $d > 0$ can strengthen stability

Paradoxically, a small complexity cost can *enhance* structural stability. Any mutant conditional strategy deviating from ISR's rule receives a Bad image and is treated like ALLD, but pays the complexity cost $d$. Hence, such mutants are never fitter than ALLD at the coexistence point. Only exact ALLD clones could invade neutrally, but ALLD is already present. Therefore, $d > 0$ prevents drift into alternative conditional strategies, leaving ISR–ALLD coexistence as the only non-defection attractor.

## S7 Complexity costs in SR

We now analyze *strong reciprocity* (SR)—downstream reciprocity plus peer punishment, *without* upstream reciprocity—under complexity costs $d \geq 0$. As in §S6, $d$ represents a per-round cognitive/implementation cost borne by the conditional strategy.

**Strong Reciprocators (SR):** Condition their actions only on partner reputation. SR players follow a pair of rules: (i) **downstream reciprocity** (cooperates with partners holding a "Nice" reputation, and (ii) **peer punishment** (punishes partners holding a "Nasty" reputation, paying cost $\gamma$ to impose penalty $\beta$).

**Reputation and assessment in SR.** We adopt a binary reputation system (Nice/Nasty) with a discriminating SR assessment rule. Roughly speaking, players whose actions are consistent with SR are judged Nice, and those that deviate from SR are judged Nasty. The exact definition is given by "multi-factor verification" as is ISR (§S2).

The expected payoffs are, in the presence of ALLD (with $y > 0$),

$$\begin{aligned} P_{SR} &= b(x + z) - cz - \gamma(x + y) - d, \\ P_{ALLD} &= bx - \beta z, \\ P_{ALLC} &= bx - c - \beta z, \end{aligned} \tag{S30-1}$$

and in the absence of ALLD (with $y = 0$),

$$\begin{aligned} P_{SR} &= b - c - d, \\ P_{ALLC} &= b - c. \end{aligned} \tag{S31-2}$$

### S7.1 On the $y$–$z$ edge (ALLD–SR) with baseline $d = 0$

On the $y$–$z$ edge ($x = 0$), the selection gradient for SR vs. ALLD is linear in $z$:

$$G(z) = P_{SR} - P_{ALLD} = (b - c + \beta + \gamma)z - \gamma. \tag{S32}$$

Hence, $G(0) = -\gamma < 0$ and $G(1) = b - c + \beta > 0$. There is a unique interior root

$$R(d) = \frac{\gamma}{b - c + \beta + \gamma} \in (0,1), \tag{S33}$$

and since $G(z)$ is increasing, $R(d)$ is a repeller along the edge. From the fact that ALLD strictly dominates ALLC, R is transversally stable against the invasion of ALLC, and thus R is a saddle in the three-strategy state space.

**Global dynamics at $d = 0$.** There exists no interior equilibrium. Interior trajectories split— they go to the 100%-ALLD state; or otherwise they go to the 100%-SR state and then drift on the x–z neutral line, eventually ending up with the 100%-ALLD state. This trajectory exemplifies an indirect invasion, where neutral drift opens the door to subsequent invasion by defectors (van Veelen, 2011).

### S7.2 On the y–z edge (ALLD–SR) with $d > 0$

Adding a complexity cost to SR shifts the intercept:

$$G_d(z) = P_{SR} - P_{ALLD} = (b - c + \beta + \gamma)z - (\gamma + d). \qquad (S34)$$

The single interior root moves to

$$R_d = \frac{\gamma + d}{b - c + \beta + \gamma}, \quad \text{exists iff} \quad d < b - c + \beta =: d_2. \qquad (S35)$$

As on the baseline, $R_d$ is a repeller along the edge (linear $G_d$ with positive slope). The effect of $d$ is to push $R_d$ toward the ALLD vertex. Then, again, $R_d$ stays transversally stable against the ALLC invasion.

**Global dynamics at $d > 0$.** The neutrality collapses, and ALLC strictly dominates SR along the x–z edge. Therefore, the SR vertex is unstable, and the evolutionary fate of the population is merely the 100%-ALLD state.

### S7.5 Contrast to ISR

SR's y–z selection gradient is linear, yielding at most one interior boundary rest point, which is always repelling along the edge (Fig. 1C). With $d > 0$, the x–z edge loses neutrality and ALLC dominates SR. The y–z interior rest point is a saddle. As a result, ALLD is the global attractor for any $d > 0$.

In sharp contrast, ISR (with upstream reciprocity) produces a quadratic y–z gradient with an unstable–stable pair of boundary equilibria (§§S4–S6) (Fig. 1B). Under small $d > 0$, ISR preserves a bistable global structure (ALLD vs. ISR–ALLD coexistence) and even gains structural stability because $d > 0$ filters out alternative conditional mutants (§S6.3). Thus, ISR—but not SR—supports productive punishment under realistic complexity costs.